\documentclass[aps,twocolumn,groupedaddress]{revtex4}
\usepackage{amssymb,amsmath}
\usepackage{graphicx,array}
\usepackage{dcolumn}
\usepackage{bm}

\begin{document}
\title{Comment on ``High-pressure phases of group-II difluorides: Polymorphism and 
       superionicity''}

\author{Claudio Cazorla}
\thanks{Corresponding Author}
\affiliation{School of Materials Science and Engineering, UNSW Sydney, Sydney NSW 2052, 
             Australia \\ Integrated Materials Design Centre, UNSW Sydney, Sydney NSW 
             2052, Australia} 

\author{Daniel Errandonea}
\affiliation{Departamento de F\'isica Aplicada (ICMUV),
             Universitat de Valencia, 46100 Burjassot, Spain}

\begin{abstract}
Nelson \emph{et al.} [Phys. Rev. B \textbf{95}, 054118 (2017)] recently have reported first-principles calculations on the behaviour of group-II difluorides (BeF$_{2}$, MgF$_{2}$, and CaF$_{2}$) under high-pressure and low- and high-temperature conditions. The calculations were based on \emph{ab initio} random structure searching and the quasi-harmonic approximation (QHA). Here, we point out that, despite the of inestimable value of such calculations at high-pressure and low-temperature conditions, the high-$P$ high-$T$ phase diagram proposed by Nelson \emph{et al.} for CaF$_{2}$ neither is in qualitative agreement with the results of previous \emph{ab initio} molecular dynamics simulations nor with the existing corps of experimental data. Therefore, we conclude that the QHA-based approach employed by Nelson \emph{et al.} cannot be applied reliably to the study of phase boundaries involving superionic phases. This conclusion is further corroborated by additional \emph{ab initio} calculations performed in the superionic compounds SrF$_{2}$, BaF$_{2}$, Li$_{3}$OCl, and AgI. 
\end{abstract}
\maketitle

\begin{figure}[t]
\centerline{
\includegraphics[width=0.90\linewidth]{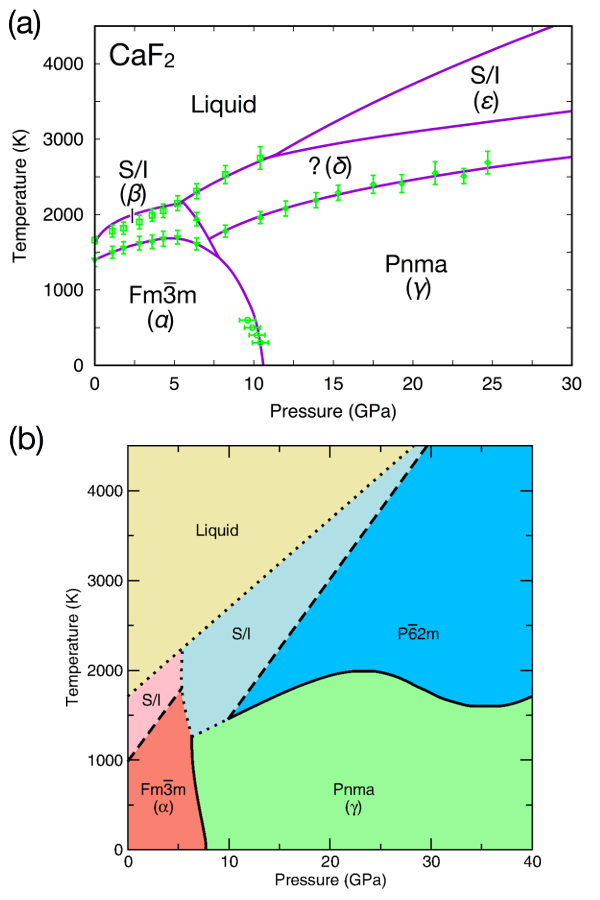}}
\caption{High-$P$ high-$T$ phase diagrams proposed for CaF$_{2}$. (a)~Adapted from work~\cite{cazorla14}; ``S/I'' stands for superionic phases and the question mark indicates that the corresponding crystal structure has not been resolved experimentally; green dots and error bars correspond to DAC experiments~\cite{cazorla14}. (b)~Adapted from work~\cite{nelson17}; the region coloured in pink with the S/I label indicates that the cubic Fm$\overline{3}$m phase develops unstable phonons; the region coloured in light blue with the S/I label indicates that the hexagonal P$\overline{6}$2m phase develops unstable phonons.}
\label{fig1}
\end{figure}

CaF$_{2}$ is an archetypal type-II fast-ion conductor in which above a particular transition temperature, $T_{s}$, the fluorine ions start to diffuse trough the crystal by successively hopping among neighbouring interstitial sites. At low-$T$ and pressures $0 \le P \lesssim 10$~GPa, CaF$_{2}$ adopts a cubic fluorite structure ($\alpha$, space group $Fm\overline{3}m$) in which the Ca$^{+2}$ cations are cubic coordinated to the F$^{-}$ anions; at low-$T$ and pressures $P \gtrsim 10$~GPa, CaF$_{2}$ presents an orthorhombic PbCl$_{2}$-like structure ($\gamma$, space group $Pnma$) in which the atomic coordination around the calcium ions is highly asymmetric. In a combined experimental and theoretical study~\cite{cazorla14}, we proposed a high-$P$ high-$T$ phase diagram for CaF$_{2}$ in which two interesting pressure-induced superionic effects were observed (see Fig.~\ref{fig1}a), namely, (i)~an anomalous decrease of $T_{s}$ in the interval $5 \lesssim P \lesssim 8$~GPa, and (ii)~a temperature-induced phase transformation from the $\gamma$ phase to an experimentally not resolved structure at $P \gtrsim 8$~GPa ($\delta$), that becomes superionic before melting ($\epsilon$ phase). Our experiments were based on diamond-anvil cell (DAC) measurements along with the laser speckle technique, and our calculations on \emph{ab initio} molecular dynamics (AIMD) simulations, which fully take into account anharmonic effects at $T \neq 0$ conditions.     

In a recent study~\cite{nelson17}, Nelson \emph{et al.} have reported first-principles calculations on the high-$P$ high-$T$ phase diagram of CaF$_{2}$ based on zero-temperature random structure searching and the quasi-harmonic approximation (QHA)~\cite{nelson17}. The authors of that study propose a new candidate structure for the high-$P$ high-$T$ $\delta$ phase with hexagonal symmetry (space group $P\overline{6}2m$, see Fig.~\ref{fig1}b); also, they estimate a series of high-$T$ coexistence lines involving superionic phases based on the QHA. The QHA strategy employed by Nelson \emph{et al.} to determine normal--superionic phase coexistence lines is as follows. Initially, the zero-temperature threshold volume at which a particular crystal structure first develops imaginary phonon frequencies, $V_{inst}$, is determined; subsequently, given a fixed pressure point one finds the temperature at which according to the QHA the volume of the system equals $V_{inst}$, namely, $V (P, T_{c}) = V_{inst}$. The series of $T_{c}$'s so obtained after considering different pressure points are then ascribed to a normal--superionic phase coexistence line. Nelson \emph{et al.} claim that their results on the phase diagram of CaF$_{2}$ are in overall qualitative agreement with those previously reported by us in work~\cite{cazorla14}.

In this Comment, we show that (i)~the high-$P$ high-$T$ CaF$_{2}$ phase diagram proposed by Nelson \emph{et al.} neither is in qualitative agreement with our previous \emph{ab initio} molecular dynamics results nor with our experiments, and (ii)~the high-$P$ high-$T$ hexagonal $P\overline{6}2m$ phase proposed by Nelson \emph{et al.} does not sustain superionicity and melts at temperatures well below the fusion line of the experimentally unresolved $\delta$ phase. Consequently, we argue that the QHA-based approach employed by Nelson \emph{et al.} in work~\cite{nelson17} is not appropriate for the study of phase boundaries involving superionic phases (as we further demonstrate by performing first-principles calculations in archetypal superionic materials other than CaF$_{2}$).      

\begin{figure}[t]
\centerline{
\includegraphics[width=0.90\linewidth]{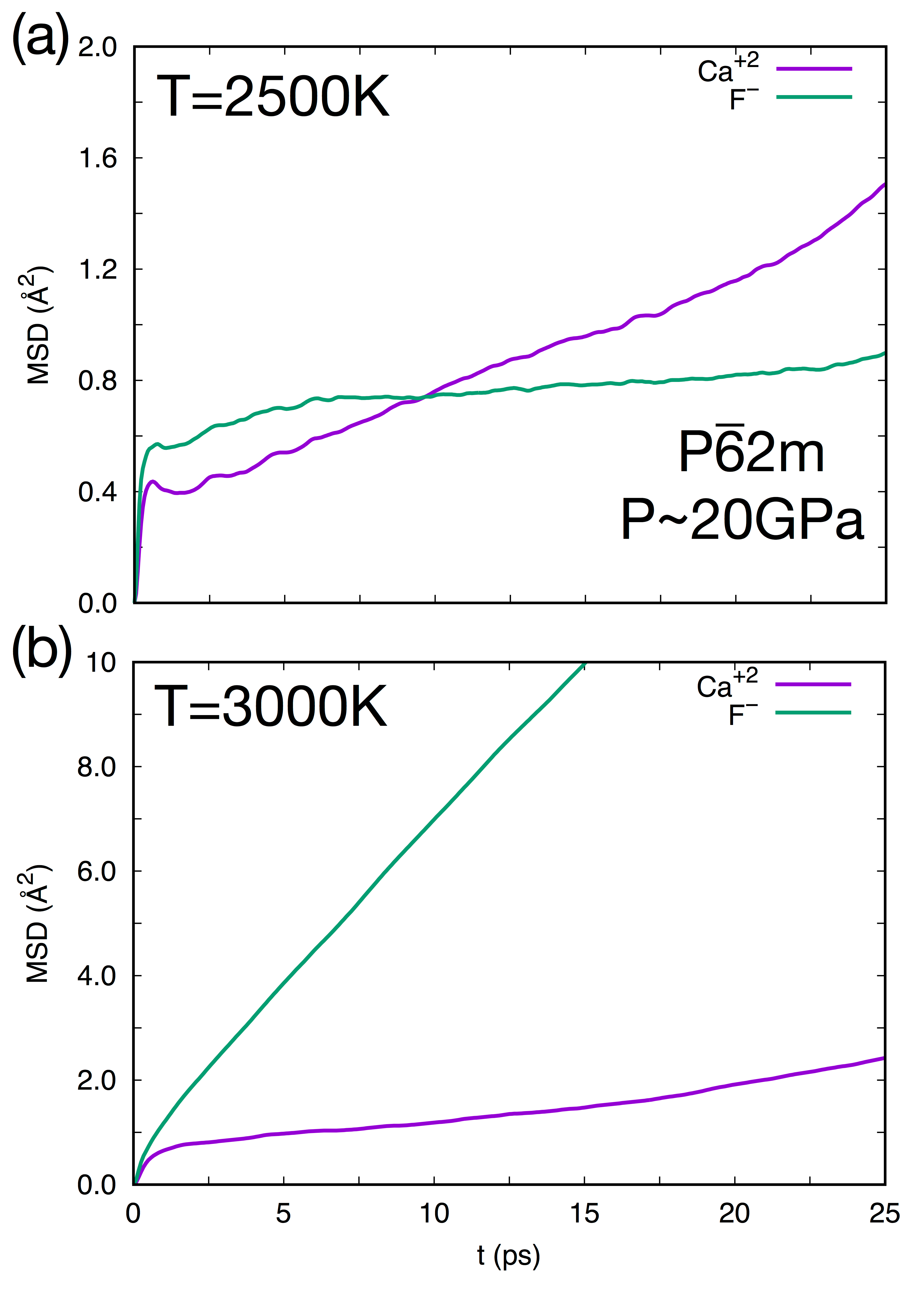}}
\caption{Ionic mean squared displacements calculated for CaF$_{2}$ in the hexagonal P$\overline{6}$2m phase at $P = 20(1)$~GPa and $2500 \le T \le 3000$~K. The slope of the curves at long simulation times are proportional to the diffusion coefficients of the ions. (a)~The sublattice of Ca$^{+2}$ ions becomes vibrationally unstable. (b)~All the ions diffuse hence the system is a liquid.}
\label{fig2}
\end{figure}

\begin{figure*}[t]
\centerline{
\includegraphics[width=0.90\linewidth]{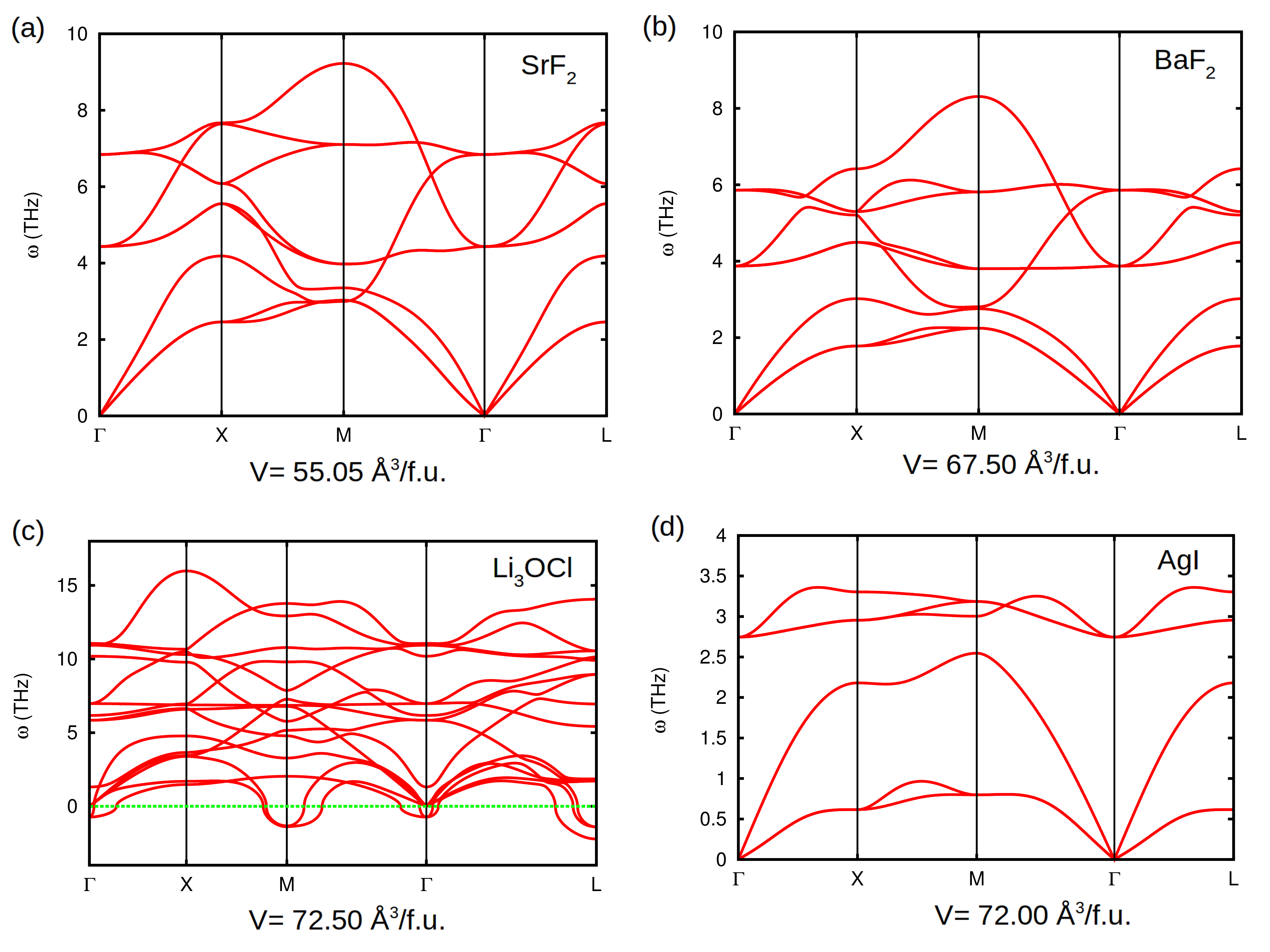}}
\caption{Vibrational phonon spectrum calculated in several archetypal superionic materials. (a)~SrF$_{2}$ considering 
         the volume at which becomes superionic at $P = 0$ in our AIMD simulations. (b)~BaF$_{2}$ considering
         the volume at which becomes superionic at $P = 0$ in our AIMD simulations. (c)~Stoichiometric Li$_{3}$OCl
         considering a volume at which imaginary phonon frequencies appear. (d)~AgI considering the volume at which 
         becomes superionic at $P = 0$ in our AIMD simulations.}
\label{fig3}
\end{figure*}

Figure~\ref{fig1} shows the high-$P$ high-$T$ phase diagrams proposed for CaF$_{2}$ in works~\cite{cazorla14} (Fig.~\ref{fig1}a) and~\cite{nelson17} (Fig.~\ref{fig1}b). A number of important quantitative and qualitative differences are obvious. First, contrary to what has been suggested by Nelson \emph{et al.}, the solid--liquid phase boundaries in both phase diagrams are not the same; in Fig.~\ref{fig1}a the slope of the solid--liquid coexistence line is not a constant. Second, Nelson \emph{et al.} propose a phase diagram in which the two superionic phases (denoted as ``S/I'' in the figure) deriving from the $\alpha$ and hexagonal $P\overline{6}2m$ structures coexist; this is in stark contrast to the results shown in Fig.~\ref{fig1}a, where coexistence between superionic phases is absent. Third, the slope of the $\gamma$--$P\overline{6}2m$ coexistence line in Fig.~\ref{fig1}b presents an unusual sign variation under increasing compression, which according to Clausius-Clapeyron implies a singular $P$-induced effect on the transition volume change (namely, $\Delta V \geq 0$ at $12 \lesssim P \lesssim 25$~GPa, $\Delta V \leq 0$ at $25 \lesssim P \lesssim 36$~GPa, and $\Delta V \geq 0$ at $P \gtrsim 36$~GPa -- mind that $\Delta S > 0$ independently of pressure --). Such a peculiar variation of the transition volume change neither is consistent with our AIMD results nor with our experimental DAC measurements reported in work~\cite{cazorla14} (see the $\gamma$--$\delta$ coexistence line and green dots in Fig.~\ref{fig1}a). Fourth, the coexistence lines involving normal and superionic phases in Fig.~\ref{fig1}b (dashed lines) invariably present constant, large, and positive slopes, as a result of the inherent limitations of the quasi-harmonic approximation; extrapolation of the S/I--$P\overline{6}2m$ coexistence line suggests the loss of superionicity at pressures above $\sim 30$~GPa. This behaviour is not consistent with the results presented in Fig.~\ref{fig1}a, in  which the equivalent phase boundaries have a less pronounced slope and the $\delta$--$\epsilon$ and solid--liquid coexistence lines do not intersect at around $30$~GPa.              

Based on \emph{ab initio} random structure searching, Nelson \emph{et al.} have proposed an hexagonal $P\overline{6}2m$ phase as the likely candidate for the $\delta$ phase appearing in Fig.~\ref{fig1}a~\cite{nelson17}. As it has been demonstrated by Nelson \emph{et al.}, and we have explicitly corrobated, that structure is energetically very competitive with respect to the $\gamma$ phase at high-$P$ and zero-temperature conditions. By employing the same AIMD techniques than in work~\cite{cazorla14}, we have analysed the superionic behaviour and structural stability of the hexagonal $P\overline{6}2m$ phase at $T \neq 0$ conditions. In Fig.~\ref{fig2}, we enclose the ionic mean-squared displacements~\cite{cazorla14} calculated at $P = 20(1)$~GPa and $2500 \leq T \leq 3000$~K. We find that at $T = 2500$~K the hexagonal $P\overline{6}2m$ phase becomes vibrationally unstable as the sublattice formed by Ca$^{+2}$ atoms \emph{melts} and the mobile cations start to diffuse through the crystal (Fig.~\ref{fig2}b); the same conclusion is reached via the computation of position correlation functions~\cite{cazorla14} (not shown here). The thermodynamic state $P = 20$~GPa and $T = 2500$~K roughly coincides with a point of the S/I--$P\overline{6}2m$ coexistence line shown in Fig.~\ref{fig1}b, however, the observed behaviour cannot be identified with superionicity as for that the F$^{-}$ anions should be diffusing instead~\cite{hull04}. At $T = 3000$~K, we find that the $P\overline{6}2m$ phase totally melts in our one-phase AIMD simulations (as all the ions are diffusing, see Fig.~\ref{fig2}c), which indicates that the corresponding melting temperature at $P = 20(1)$~GPa is very likely to lie below that point~\cite{cazorla12,cazorla08,bouchet09}. Since the melting temperature that we accurately calculated for the $\epsilon$ phase at $P = 20(1)$~GPa by means of two-phase coexistence AIMD simulations is noticeably above $3000$~K (see Fig.~\ref{fig1}a), we may conclude that the hexagonal $P\overline{6}2m$ structure is not a good candidate for either the $\delta$ or $\epsilon$ phases proposed for CaF$_{2}$ in work~\cite{cazorla14}. Likewise, we conclude that the QHA-based approach employed by Nelson \emph{et al.} is not appropriate for describing superionic CaF$_{2}$ at high temperatures, as due to the neglection of anharmonic effects that are inherent to fast-ion conductors (e.g., the creation of $T$-induced lattice defects).           
  
To further assess the performance of the QHA-based method introduced by Nelson \emph{et al.} in identifying normal--superionic transition points in general, 
we have performed additional phonon calculations and AIMD simulations in the fast-ion conductors SrF$_{2}$, BaF$_{2}$, Li$_{3}$OCl, and AgI. 
For SrF$_{2}$ and BaF$_{2}$, the zero-pressure normal--superionic transition temperatures that we have estimated with AIMD 
simulations are $T_{s}(0) = 1150(100)$ and $1135(100)$~K, respectively, which are in very good agreement with the available experimental 
data~\cite{voronin01,evangelakis87}. In Figs.~\ref{fig3}a-b, we show the phonon spectra calculated at the volumes corresponding 
to those superionic transition points; as can be appreciated therein no imaginary phonon frequencies develop. For Li$_{3}$OCl, 
we have first calculated the threshold volume at which imaginary phonon frequencies begin to appear (see Fig.~\ref{fig3}c). 
Then, by constraining that volume, we have performed a series of AIMD simulations in which the temperature is steadily raised 
until reaching a completely melt state ($T \le 1500$~K). We have observed that at such conditions the system never becomes 
superionic; the reason for this result is that, as it is well known, superionicity only appears in non-stoichiometric Li$_{3}$OCl 
systems~\cite{zhang13,lu15,cazorla17}. Finally, we have analyzed AgI where the normal--superionic phase transition is of 
first-order type as it has associated a large latent heat, change of volume, and the crystal symmetry of both the cation and 
anion sublattices changes during the transformation~\cite{hull04,cazorla17,parrinello83}. Again, the phonon spectrum calculated 
at the volume at which the system becomes superionic at zero pressure ($T_{s}(0) = 400(20)$~K~\cite{hull04}) does not exhibit 
any imaginary phonon frequency (see Fig.~\ref{fig3}d). We note that in this latter case the QHA could have been expected to 
be valid, as the superionic transition temperature is well below the corresponding melting temperature, $T_{m}(0) = 840(20)$~K; 
however, due to the first-order character of the transformation, superionicity neither can be identified through the analysis 
of unstable phonon modes.  

In conclusion, we have shown that the QHA-based approach introduced by Nelson \emph{et al.} in work~\cite{nelson17} is not appropriate 
to describe phase coexistence lines involving superionic phases in the high-$P$ high-$T$ phase diagram of CaF$_{2}$, and in general in 
any superionic material. The main reasons behind such a QHA failure are the neglection of $T$-induced anharmonic effects, like the creation 
of lattice defects, which are crucial for the stabilization of the superionic state, and the fact that the normal--superionic phase 
transition normally is not soft-phonon mode driven (as exemplified by Li$_{3}$OCl and AgI). Our criticisms on the QHA approach used by
Nelson \emph{et al.} to determine superionic transition points appear to be backed also by the experimental evidence, as superionicity 
occurs only in very specific compounds~\cite{hull04} whereas the vast majority of materials present positive thermal expansions and zero-temperature 
threshold volumes at which imaginary phonon frequencies develop. In addition, we have shown that the hexagonal $P\overline{6}2m$ 
phase proposed by Nelson \emph{et al.} is neither superionic nor vibrationally stable at high-$P$ high-$T$ conditions. 

This research was supported under the Australian Research Council's Future Fellowship funding
scheme (No. FT140100135), the Spanish government MINECO, the Spanish Agencia Estatal de Investigacion 
(AEI), and the Fondo Europeo de Desarrollo Regional (FEDER) (No. MAT2016-75586-C4-1-P and 
MAT2015-71070-REDC). Computational resources and technical assistance were provided by the Australian 
Government and the Government of Western Australia through Magnus under the National Computational 
Merit Allocation Scheme and The Pawsey Supercomputing Centre.

\end{document}